\documentclass[letterpaper, 10 pt, conference]{ieeeconf}
\IEEEoverridecommandlockouts
\usepackage{acro}
\let\labelindent\relax
\usepackage{enumitem}

\usepackage{graphics}
\usepackage{graphicx} 
\usepackage{multirow}
\usepackage{longtable}
\usepackage{epsfig}
\usepackage{epstopdf}

\usepackage{amsmath, mathrsfs, dsfont}
\usepackage{amssymb}
\usepackage{amsfonts}
\usepackage{bm}

\usepackage{cite}
\usepackage{verbatim}

\usepackage{amscd}
\usepackage{color}
\definecolor{lgray}{gray}{0.6}
\usepackage{rotating}

\usepackage[font=footnotesize]{subcaption}

\usepackage{mathptmx}
\usepackage[11pt]{moresize}
\usepackage{flushend}
\usepackage[stable]{footmisc}

\usepackage{algorithmicx}
\usepackage{algorithm}
\usepackage{algpascal}
\usepackage{float}
\usepackage{algc}
\usepackage{algcompatible}
\usepackage{algpseudocode}

\usepackage[vcentermath]{youngtab} 
\usepackage{url}

\newcommand{\BoldP}				{ \mathbf{P} }

\newcommand{\BoldR}				{ \mathbf{R} }

\newcommand{\0}					{ \boldsymbol{0} }

\newcounter{inlineenum}
\renewcommand{\theinlineenum}{\alph{inlineenum}}

	\newcommand{\brmrk}[1]{\begin{remark} \label{#1} }
	\newcommand{\ermrk}{ \hfill $\bigtriangleup$    \end{remark} \vspace{1mm} }

\newtheorem{exercise}{Exercise}[section]
\newcommand{\boex}[1]{\begin{example} \label{#1} --- \rm}
	\newcommand{\eoex}{ \hfill $\bigtriangleup$    \end{example} \vspace{1mm} }
\newtheorem{example}{Example}[section]
\newcommand{\bohw}[1]{\begin{exercise} \label{#1} -- \rm}
	\newcommand{\eohw}{ \hfill    \end{exercise} \vspace{1mm} }
\newtheorem{assumption}{Assumption}[section]
	\newcommand{\boass}[1]{\begin{assumption} \label{#1} -- \rm}
	\newcommand{\eoass}{ \hfill    \end{assumption} \vspace{1mm} }

\renewcommand{\baselinestretch}{1.0}

\newcommand{\black}{\color{black}}

\definecolor{brinkpink}{rgb}{1.00, 0.33, 0.64}



\DeclareAcronym{APC}{
	short=APC,
	long=Antenna Phase Centre,
}

\DeclareAcronym{COM}{
short=COM,
long=Center of Mass,
}

\DeclareAcronym{CV}{
short=CV,
long=Connected Vehicle,
}

\DeclareAcronym{CAV}{
	short=CAV,
	long=Connected and Automated Vehicle,
}

\DeclareAcronym{DCB}{
short=DCB,
long=Differential Code Bias,
}

\DeclareAcronym{DSRC}{
	short=DSRC,
	long=Dedicated Short-Range Communications,
}

\DeclareAcronym{CDMA}{
	short=CDMA,
	long=Code Dvision Multiple Access,
}

\DeclareAcronym{CDF}{
	short=CDF,
	long=Cumulative Distribution Function,
}

\DeclareAcronym{CE-CERT}{
	short=CE-CERT,
	long=College of Engineering Center for Environmental Research and Technology,
}

\DeclareAcronym{FDMA}{
	short=FDMA,
	long=Frequency Dvision Multiple Access,
}
\DeclareAcronym{DGPS}{
	short=DGPS,
	long=Differential Global Positioning System,
}
\DeclareAcronym{DGNSS}{
short=DGNSS,
long=Differential GNSS,
}
\DeclareAcronym{ECEF}{
short=ECEF,
long=Earth-Centered Earth-Fixed,
}
\DeclareAcronym{GPS}{
	short=GPS,
	long=Global Positioning System,
}
\DeclareAcronym{GNSS}{
short=GNSS,
long=Global Navigation Satellite Systems,
}
\DeclareAcronym{GPSSPS}{
short=GPS SPS,
long=GPS standard positioning service,
}

\DeclareAcronym{HMM}
{short = HMM, long = Hidden Markov Model}

\DeclareAcronym{HD}
{short = HD, long = Horizontal Distance}

\DeclareAcronym{HiD}{
	short=Hi-Def,
	long=High-definition,
}

\DeclareAcronym{IMU}{
short=IMU,
long=Inertial Measurement Unit,
}

\DeclareAcronym{I2V}{
short=I2V,
long=Infrastructure-to-Vehicle,
}

\DeclareAcronym{LLMM}
{short = LLMM, long = Lane-level Map-matching}

\DeclareAcronym{RLMM}
{short = RLMM, long = Road-level Map-matching}

\DeclareAcronym{MSE}
{short = MSE, long = Mean Square Error}

\DeclareAcronym{OS}{
	short=OS,
	long=Open Service,
}

\DeclareAcronym{OSB}{
short=OSB,
long=Observable-specific Code Biases,
}
\DeclareAcronym{OSR}{
short=OSR,
long=Observation Space Representation,
}
\DeclareAcronym{PPP}{
	short=PPP,
	long=Precise Point Positioning,
} 
\DeclareAcronym{PPP-AR}{
short=PPP-AR,
long=Precise Point Positioning Ambiguity Resolution,
} 
\DeclareAcronym{RTCM}{
short=RTCM,
long=Radio Technical Commission for Maritime Services,
}
\DeclareAcronym{RTK}{
short=RTK,
long=Real-time Kinematic Positioning,
}
\DeclareAcronym{SBAS}{
short=SBAS,
long=Satellite Based Augmentation Systems,
}

\DeclareAcronym{SNR}{
short=SNR,
long=Signal-to-Noise Ratio,
}
\DeclareAcronym{SSR}{
short=SSR,
long=State Space Representation,
}
\DeclareAcronym{SPS}{
	short=SPS,
	long=Standard Positioning Service,
}

\DeclareAcronym{SAE}{
	short=SAE,
	long=Society of Automotive Engineers,
}
\DeclareAcronym{STEC}{
	short=STEC,
	long=Slant Total Electron Content,
}
\DeclareAcronym{VTEC}{
	short=VTEC,
	long=Vertical Total Electron Content,
}

\DeclareAcronym{SH}{
	short=SH,
	long=Spherical Harmonic,
}
\DeclareAcronym{TGD}{
short=TGD,
long=Timing Goup Delay,
}
\DeclareAcronym{ZTD}{
short=ZTD,
long=Zenith Troposphere Delay,
}
\DeclareAcronym{TEC}{
short=TEC,
long=Total Electron Content,
}
\DeclareAcronym{IPP}{
	short=IPP,
	long=Ionosphere Pierce Point,
}
\DeclareAcronym{NOAA}{
	short=NOAA,
	long=National Oceanic and Atmospheric Administration,
}
\DeclareAcronym{UCR}{
	short=UCR,
	long=University of California-Riverside,
}
\DeclareAcronym{USTEC}{
short=US-TEC,
long=US Total Electron Content,
}
\DeclareAcronym{VNDGNSS}{
	short=VN-DGNSS,
	long=Virtual Network DGNSS,
}
\DeclareAcronym{IOD}{
	short=IOD,
	long=Issue Of Data,
}
\DeclareAcronym{CAS}{
	short=CAS,
	long=Chinese Academy of Sciences,
}
\DeclareAcronym{CNES}{
	short=CNES,
	long=Centre national d'études spatiales,
}
\DeclareAcronym{RMS}{
	short=RMS,
	long=Root Mean Square,
}
\DeclareAcronym{SF}{
	short=SF,
	long=Single Frequency,
}
\DeclareAcronym{STD}{
	short=STD,
	long=Standard Deviation,
}
\DeclareAcronym{DF}{
	short=DF,
	long=Dual Frequency,
}
\DeclareAcronym{ICD}{
	short=ICD,
	long=Interface Control Document,
}
\DeclareAcronym{NED}{
	short=NED,
	long={North, East and Down},
}
\DeclareAcronym{WAAS}{
	short=WAAS,
	long=Wide Area Augmentation System,
}
\DeclareAcronym{OPUS}{
	short=OPUS,
	long=Online Positioning User Service,
}

\DeclareAcronym{VRS}{
	short=VRS,
	long=Virtual Reference Station,
}

\DeclareAcronym{SPP}{
	short=SPP,
	long=Single-frequency Point Positioning,
}

\DeclareAcronym{SPaT}{
	short=SPaT,
	long=Signal Phase and Timing,
}

\DeclareAcronym{BNC}{
	short=BNC,
	long=BKG NTRIP Client,
}

\DeclareAcronym{ITS}{
	short=ITS,
	long=Intelligent Transportation System
}

\DeclareAcronym{USDOT}{
	short=USDOT,
	long=U.S. Department of Transportation
} 

\begin{document}
	\date{\today}
	\title{Assessment of U.S. Department of Transportation \\ Lane-Level Map for Connected Vehicle Applications}
	\author{Wang~Hu,
		David~Oswald,
		Guoyuan~Wu,~\IEEEmembership{Senior Member,~IEEE}
		and~Jay~A.~Farrell,~\IEEEmembership{Fellow,~IEEE}
		\thanks{This research was conducted as part of the ``Lane-Level Localization and Map Matching for Advanced Connected and Automated Vehicle (CAV) Applications'' project funded by the National Center for Sustainable Transportation (NCST) during the period of performance from April 1, 2021 through March 31, 2022. }
	}
	\maketitle
	\begin{abstract}
		\ac{HiD} digital maps are an indispensable  automated driving technology that is developing rapidly. 
		There are various commercial or governmental map products in the market. 
		It is notable that the \ac{USDOT} map tool allows the user to create MAP and \ac{SPaT} messages with free access. 
		However,  an analysis of the accuracy of this map tool is currently lacking in the literature.	
		This paper provides such an analysis. 
		The analysis manually selects 39 feature points within about 200 meters of the verified point and 55 feature points over longer distances from the verified point. 
		All feature locations are surveyed using GNSS and mapped using the USDOT tool. 
		Different error sources are evaluated to allow assessment of the USDOT map accuracy.  
		In this investigation, The USDOT map tool is demonstrated to achieve 17 centimeters horizontal accuracy, which meets the lane-level map  requirement. The maximum horizontal  map error is less than 30 centimeters.
		
	\end{abstract}
		
	\section{Introduction}
	
Digital roadway maps have a long history, especially at the roadway-level, providing information about road inter-connectivity with positions accurate to the decimeter level \cite{white1991}. Using sensors such as LiDAR, radar, and digital cameras, along with techniques such as Geographic Information System (GIS) and Machine Learning, the precision of roadway maps can reach centimeter level \cite{massow2016, jiao2018}.

The advent of automated vehicles has motivated interest in \ac{HiD} digital maps which may include different capabilities: road-level, lane-level, and road features \cite{liu2020}. \ac{HiD} digital maps are a fundamental technology for  connected and cooperative vehicles  enabling applications such as:
GNSS-based lane recognition \cite{rogers1999, yan2020},
per lane queue determination \cite{poto2017},
per lane over-speed warning \cite{logo2021},  etc. 
General Motors Co. had every mile of interstate in the United States and Canada mapped using LiDAR for the Cadillac’s Super Cruise, which is a hands-off semi-autonomous system \cite{pal2019}. Waymo, Uber Technologies, and Ford Motor Co. also have fleets of vehicles out to create \ac{HiD} maps for use in autonomous vehicles \cite{szanto2019}. Mobileye collects data from the millions of customers’ vehicles using their front-facing cameras, and combines the data to create \ac{HiD} maps that will have road features, e.g., lane markers, traffic signals, and road boundaries \cite{abrams2017, wong2020}. Mobileye’s approach allows for maps to be constantly updated for temporary features such as constructions and roadblocks \cite{wong2020}.

Other companies have adopted Mobileye’s crowd-sourcing approach to create \ac{HiD} maps: TomTom (tomtom.com)
and 
HELLA Aglaia (hella-aglaia.com)
partnered to create crowd-sourced \ac{HiD} maps; NVIDIA
\footnote{URL:\url{https://www.nvidia.com/en-us/self-driving-cars/hd-mapping/}}
uses camera and radar data from their autonomous vehicle platform; Bosch and Volkswagen\footnote{URL: \url{https://www.bosch-presse.de/pressportal/de/en/swarm-intelligence-for-automated-driving-231431.html}} partnered to create \ac{HiD} maps by leveraging data from sensors and a digital twin model; and Mitsubishi teamed up with Woven Planet (woven-planet.global) to use their Automated Mapping Platform that uses vehicle sensor data and satellite imagery to create \ac{HiD} maps.

While companies are working on automated technologies for high-definition roadway maps with global extent and commercial tools are available for manual construction of such maps for smaller regions, currently there are few free tools available to researchers for projects and demonstrations. 

The notable exception is the \ac{USDOT} J2735 MAP tool, discussed in Section \ref{sec:tool}. 
It provides a web application user interface that uses satellite imagery to enable users to manually select and map  lanes and features to create J2735 MAP messages. 
J2735 MAP messages describe an intersection’s physical layout, such as lanes, stop bars, and allowed maneuvers, in a digital form standardized by the \ac{SAE} \cite{hedges2008overview}. 
The \ac{USDOT} MAP tool can generate binary outputs as specified for \ac{DSRC} roadside units \cite{5888501} or usable through cellular \ac{I2V} communications. 
At present, the literature does not include any assessment of the accuracy of the maps produced by the \ac{USDOT} map tool. 
In addition, the establishment of MAP or \ac{SPaT} message in this map tool requires a verified point for each intersection. 
The assessment of the effective range of one verified point is conducive to decrease the demand of the amount of verified points for dense MAP messages.

This paper investigates the accuracy of the \ac{USDOT} map tool by comparing with the feature locations determined within that tool with the feature locations determined by \ac{GNSS} survey. 
Section \ref{sec:tool}  introduces the USDOT tool.
Section \ref{sec:method} presents the data acquisition methods using both \ac{GNSS} survey and the \ac{USDOT} map tool. 
It also defines and quantifies the various error sources involved in the analysis. 
Section \ref{sec:acc} assesses the overall map accuracy and discusses the bias induced by any inaccuracy of the verified point.

	\section{USDOT Connected Vehicles Tool} \label{sec:tool}      
	 The USDOT Connected Vehicles Tool (available at \url{https://webapp2.connectedvcs.com/}) offers free on-line access to tools for creating maps to support various  \ac{CAV} message types.
The {\em ISD Message Creator} constructs lane-level intersection maps to support MAP and \ac{SPaT} messages.
The detailed instructions under the ``Help'' button make the site self-explanatory. 
Our interest herein is assessing the position accuracy of  the J2735 MapData message output by the tool.

	\section{Accuracy Assessment Method} \label{sec:method}
	Accuracy will be assessed by comparing the coordinates of feature points determined by two different methods: the US-DOT Map Tool and \ac{GNSS} \ac{RTK} survey. 
These feature points are selected to satisfy the following specifications:
\begin{itemize}
	\item Each point should be easily and uniquely identifiable both to the surveyor and within the USDOT tool.
	This is typically achieved by defining the points to be at the intersection of two nearly orthogonal lines.	
	\item Each point should have a clear view of the sky.
	\item Each point should be near, but not on  the road.
	This constraint is added to ensure the safety of the person performing the survey without needing to interrupt normal traffic operation. 
	
	\item The features in the USDOT imagery and the real environment should be at the same locations. 
	The USDOT imagery is based on georectification of historic photos that may have been taken months in the past; therefore, recent changes in the real environment may not be accurately represented in that imagery. 

\end{itemize}
The ESRI  manual\footnote{See 
\url{https://pro.arcgis.com/en/pro-app/latest/help/data/imagery/overview-of-georeferencing.htm}.}
recommends using at least three points to construct a georeference. 
Topan and Kutoglu \cite{topan2009geo} use approximately thirty points for evaluation of  georeferencing accuracy. 
A typical J2735 intersection map extends for a few hundred meters along each intersection approach.
The ninety-four feature points evaluated herein extend for over 9000 m from the intersection verified point.
Therefore, the feature points used herein well exceed both the number and geographic extent  of georeferencing points  necessary for evaluation of the USDOT Map Tool for the purpose of constructing a J2735 MAP message for a typical intersection intersection.

\begin{figure*}[t]
	\centering
	\includegraphics[width=0.9\textwidth]{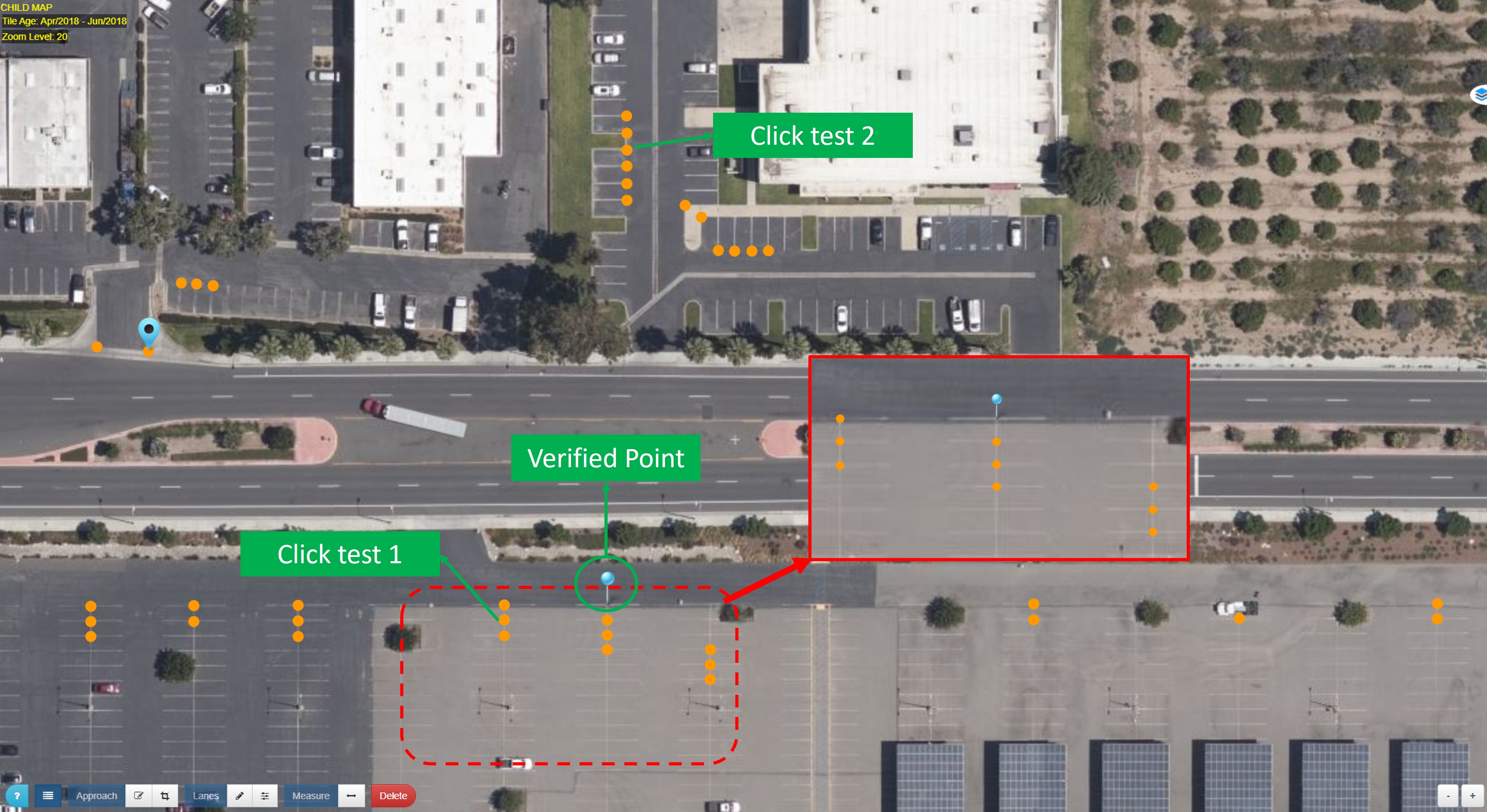}
	\caption{USDOT map accuracy test points near \ac{UCR} \ac{CE-CERT}}
	\label{fig:map1}
\end{figure*}

\begin{figure}[t]
	\centering
	\includegraphics[width=0.75\linewidth]{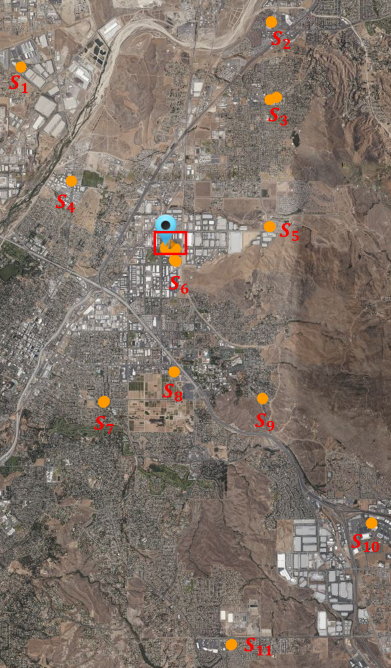}
	\caption{Expanded test area for accuracy analysis.}
	\label{fig:map2}
\end{figure}

Figures \ref{fig:map1} and \ref{fig:map2} use imagery from the USDOT tool to show the geographic distribution of the feature points.
Each orange dot in Fig. \ref{fig:map1} shows the location of each of $N_1=39$ feature points near \ac{UCR} \ac{CE-CERT}.
One of these points is defined as the verified point (denoted  $\BoldP_v^e$) 
for the USDOT tool.
The feature points in Fig. \ref{fig:map1} are each within about 200 meters of the verified point. 
The solid red box displays the region within the dashed red line at the maximum zoom level allowed by the tool.
Fig. \ref{fig:map2} uses orange dots to indicate the location of 11 survey areas. 
Each survey area, labeled from $S_1$ to $S_{11}$, includes 5 feature points. 
These $N_2=55$ feature points allow accuracy to be assessed over longer distances from the verified point.
The red box in Fig. \ref{fig:map2} indicates the region portrayed in Fig. \ref{fig:map1}.

To assess accuracy, we compare GNSS survey and USDOT mapping tool locations for each feature point. 
The symbol $\BoldP^e$ denotes the feature position determined by GNSS survey. The symbol $\hat{\BoldP}^e$ denotes the  position of the feature point determined by the USDOT mapping tool.
The superscript on the vector $\BoldP$ denotes the frame-of-reference, such as $e$ for \ac{ECEF} and $g$ for the \ac{NED} frame. 
The NED frame feature location of a point ${\BoldP}^g$ is computed by
\begin{equation}
	{\BoldP}^g = \BoldR^g_e\,({\BoldP}^e-\BoldP^e_v)
	\label{eqn:Transl_E2G}
\end{equation}
where $\BoldP^e_v$ is the origin of the NED frame and 
$\BoldR^g_e$ is the rotation matrix from the \ac{ECEF} frame to the  \ac{NED} frame \cite{hu2021using}. 
Eqn. (\ref{eqn:Transl_E2G}) is valid both for GNSS survey and USDOT mapping tool locations. 

Section \ref{sec:survey} discusses the GNSS survey and assesses its sources of error when determining the coordinates of each point.
Section \ref{sec:mark} discusses the USDOT tool and assesses its sources of error when determining the coordinates of each point.
Section \ref{sec:acc} combines the USDOT and GNSS data to assess the overall map accuracy.

\subsection{Data Acquisition: GNSS Survey}
\label{sec:survey}

This section presents the procedure for determining the real-world position of the verified point and of each feature point (denoted  $\BoldP_k^e$ for $k=1,\ldots,N$) by use of \ac{GNSS} \ac{RTK} survey, using a  dual-frequency ublox ZED-F9P receiver connected to a dual-band ublox antenna.
The antenna is placed on the ground above the corresponding feature point. 
The receiver communicates with the \ac{UCR} base station to obtain \ac{RTCM} corrections and reports \ac{RTK} fixed position solution in WGS84 \ac{ECEF} frame.

During the survey process for each point, the ZED-F9P in \ac{RTK} fixed mode was used to record the position for at least 20 seconds. 
The mean of these measurements is used in Sec. \ref{sec:acc} as the surveyed position.
The standard deviation of each coordinate in each surveyed position is less than 0.005 m.
The \ac{RTK} \ac{GNSS} surveyed position \ac{MSE}, denoted herein as $\sigma_G$ at the centimeter level (see e.g., Table 21.7 in \cite{teunissen2017springer}).

In addition to the \ac{RTK} \ac{GNSS} survey position error characterized by  $\sigma_G$, there is also antenna placement error due to the  fact that the human operator cannot perfectly place the antenna over the feature and account for the antenna phase offset.
This error is accounted for by the symbol $\sigma_S$ with \ac{MSE}  $\sigma_S=0.01\, m$.

\subsection{Data Acquisition: USDOT Map Tool}\label{sec:mark}

The goal of this section is two-fold: (1) to describe the process by which this tool was used to obtain the geodetic coordinates for the selected locations; and (2) to define and assess the related sources of error. 

{\bf Process. } Starting from the URL for the USDOT tool given in Section \ref{sec:tool}, the steps are as follows: 
\begin{enumerate}
		\item In the ISD Message Creator, 
		\begin{enumerate}
			\item Click `View Tool', then under `File' button click `New Parent Map'.
			\item Center the map imagery over the region of interest at the `Zoom Level 21', which is the highest resolution, as shown in the inset of Fig. \ref{fig:map1}.  
			\item 	Click `Builder' from the left bottom corner. 
			\item Drag the 
			`Verified Point Marker' to the  feature point defined in Section \ref{sec:method} and shown in Fig. \ref{fig:map1}. 
			A `Verified Point Configuration' window will automatically open. Input the GNSS survey coordinates for the verified Latitude/Longitude/Elevation. 	\label{err:click_verified}
			\item Drag the `Reference Point Marker' near the verified point in the map. 
			The reference point is required for the tool. 
			 It determines the relative position of all feature locations in the J2735 map message, but does not affect the results of the experiments. 
		\end{enumerate}

	\item Under the `File' button from the top menu, select `New Child Map'. 
	Click `Cancel' for the popup questions. 
	Use the pencil in the `Lanes' button located near the left bottom corner.
	Double-click each desired feature location.
	An orange dot will be displayed as shown Fig. \ref{fig:map1}.
	\label{err:click_feature}

	\item  Click the pencil in the `Lanes' button to turn it off. 
	Then, select (i.e., mouse click) each feature point in the tool imagery (e.g., orange points in Fig. \ref{fig:map1}) and note their coordinates as $\hat{\BoldP}^e_k$.
\end{enumerate}
Note that all positions acquired from the USDOT tool are WGS84 ECEF geodetic coordinates.

{\bf Error Sources. } 
The above process allows measurement error to occur in at least two ways.
First, the user will have error in the clicking of points.
For example, Steps \ref{err:click_verified} and \ref{err:click_feature} involve mouse clicks to select points. 
At best, the accuracy of such mouse clicks will be the size of the pixel in meters; however, the screen resolution may result in lower accuracy.
The click error will be denoted by 
$\sigma_C$.
Second, the geodetic coordinates assigned to the clicked points will be imperfect due to georectification errors. 
This mapping error will be denoted by 
$\sigma_M$.

{\bf Error Assessment. }
The goal of this subsection is to characterize the click accuracy $\sigma_C$ in meters. 
Point-click experiments are performed for two feature points, 
which are marked as `Click test' in Fig. \ref{fig:map1}. 
For each experiment, using `Zoom level 21', the targeted feature point is manually clicked 15 times (moving the cursor away and returning it between clicks) and their position $\hat{\BoldP}^e_{C_i}$ is recorded  for $i = 1,...,15 $.
\black

The accuracy analysis is performed in a locally-level \ac{NED} tangent frame with its origin point at the verified position $\hat{\BoldP}^e_v$. 
The NED  feature location $\hat{\BoldP}^g_{C_i}$ is computed from $\hat{\BoldP}^e_{C_i}$ using Eqn. (\ref{eqn:Transl_E2G}).

Herein, click test error is characterized by the \ac{STD} of each component of ${\BoldP}^g_{C_i}= [ P^g_{N_i},  P^g_{E_i},  P^g_{D_i}]^T$.
The \ac{STD} of North $\sigma_N$ and East $\sigma_E$ are listed in Table \ref{tab:std_click}. 
The vertical STD $\sigma_V$ is 0 since there are no changes in the Down coordinates in each click of each experiment.
The horizontal STD, which defines the click accuracy $\sigma_C$, is calculated by
\begin{equation}
	\sigma_C = \sqrt{\sigma_N^2+\sigma_E^2}.
\end{equation}
The values of  $\sigma_C$ summarized in Table \ref{tab:std_click}, will be used in Section \ref{sect:ass_map} to estimate a value for $\sigma_M$.

\begin{table}[]
	\centering
	\caption{Standard deviation for click test}
	\begin{tabular}{|c|c|c|c|c|}
		\hline
		& $\sigma_N$ & $\sigma_E$  & $\sigma_C$ & $\sigma_V$ \\ \hline
		Click test 1 & 0.053 m                                        & 
		0.028 m                                                    & 
		0.060 m                                                          & 
		0.0 m                                                            \\ \hline
		Click test 2 & 0.042 m                                                     & 
		0.032 m                                                    & 
		0.053 m                                                          & 
		0.0 m                                                            \\ \hline
	\end{tabular}
	\label{tab:std_click}
\end{table}

	\section{Accuracy Assessment} \label{sec:acc}
	This section uses the USDOT data in comparison with the GNSS survey data to assess the accuracy of the feature locations provided by the USDOT mapping tool.

\subsection{Bias Analysis on USDOT Tool: Verified Point}
\begin{figure}[tb]
	\centering
	\includegraphics[width=0.8\linewidth]{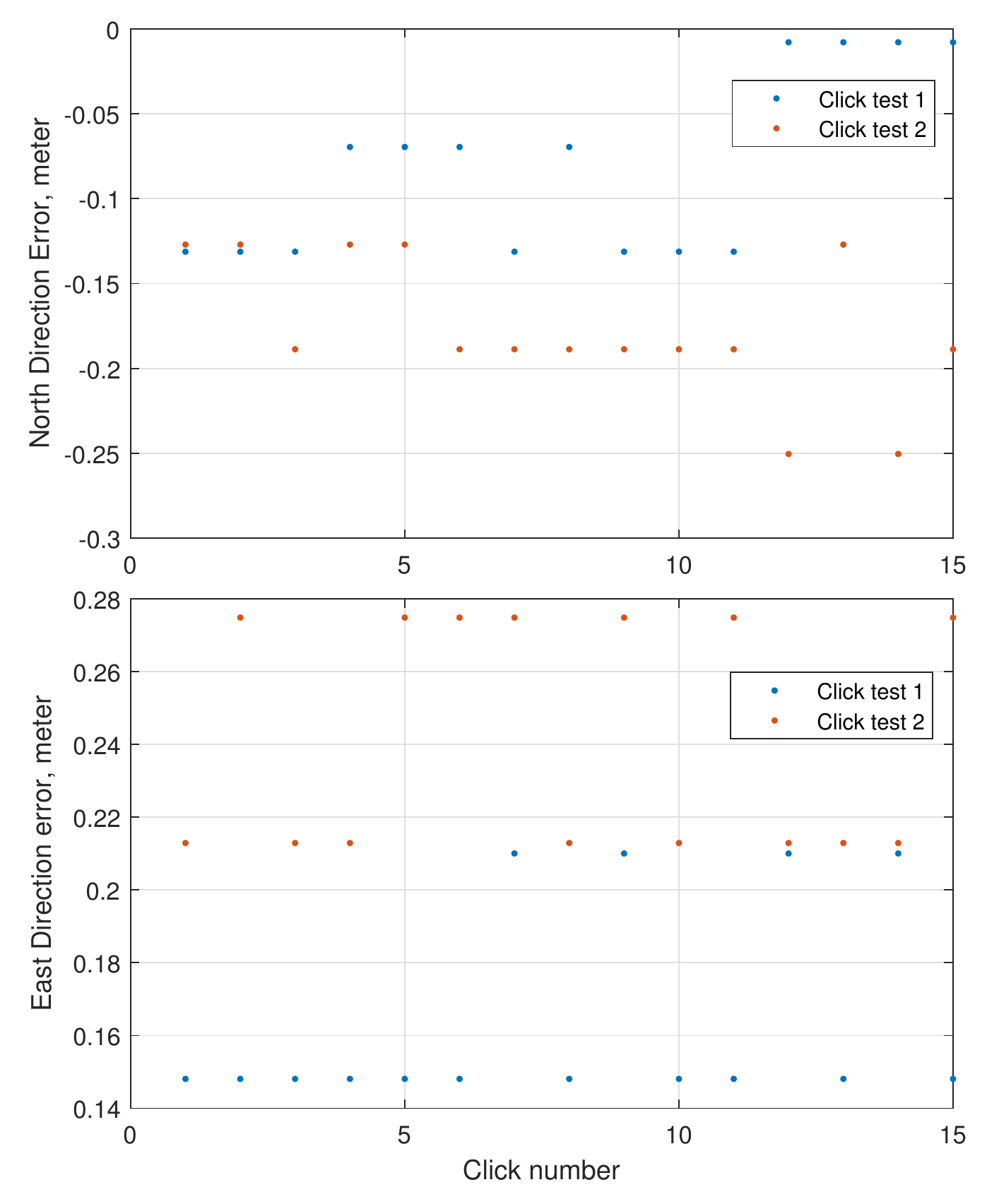}
	\caption{North and east errors between USDOT tool and GNSS survey (i.e, $\delta P^g_{N_i}$ and $\delta P^g_{E_i}$ from eqn. (\ref{eqn:geographic_err})) using the two feature points from the click test.}
	\label{fig:click}
\end{figure}

Fig. \ref{fig:click} displays the north and east components of the error between the USDOT feature points from each click test $\hat{\BoldP}^g_{C_i}$ and the GNSS surveyed positions ${\BoldP}^g_{C}$ for the same features. 
For each click test, the NED frame positions  $\hat{\BoldP}^g_{C_i}$ and ${\BoldP}^g_{C}$ are computed using Eqn. (\ref{eqn:Transl_E2G}).
The position error is calculated by
\begin{equation}\label{eqn:geographic_err}
	\delta{\BoldP}^g_{C_i}  = {\BoldP}^g_{C_i} -  {\BoldP}^g_{C}
\end{equation}
where $\delta{\BoldP}^g_{C_i}=[\delta P^g_{N_i}, \delta P^g_{E_i}, \delta P^g_{D_i}]^T$ are the NED components of mapping error for click test $C_i$.

\begin{figure*}[tb]
	\centering
	\begin{subfigure}{.4\textwidth}
		\includegraphics[width=0.9\linewidth]{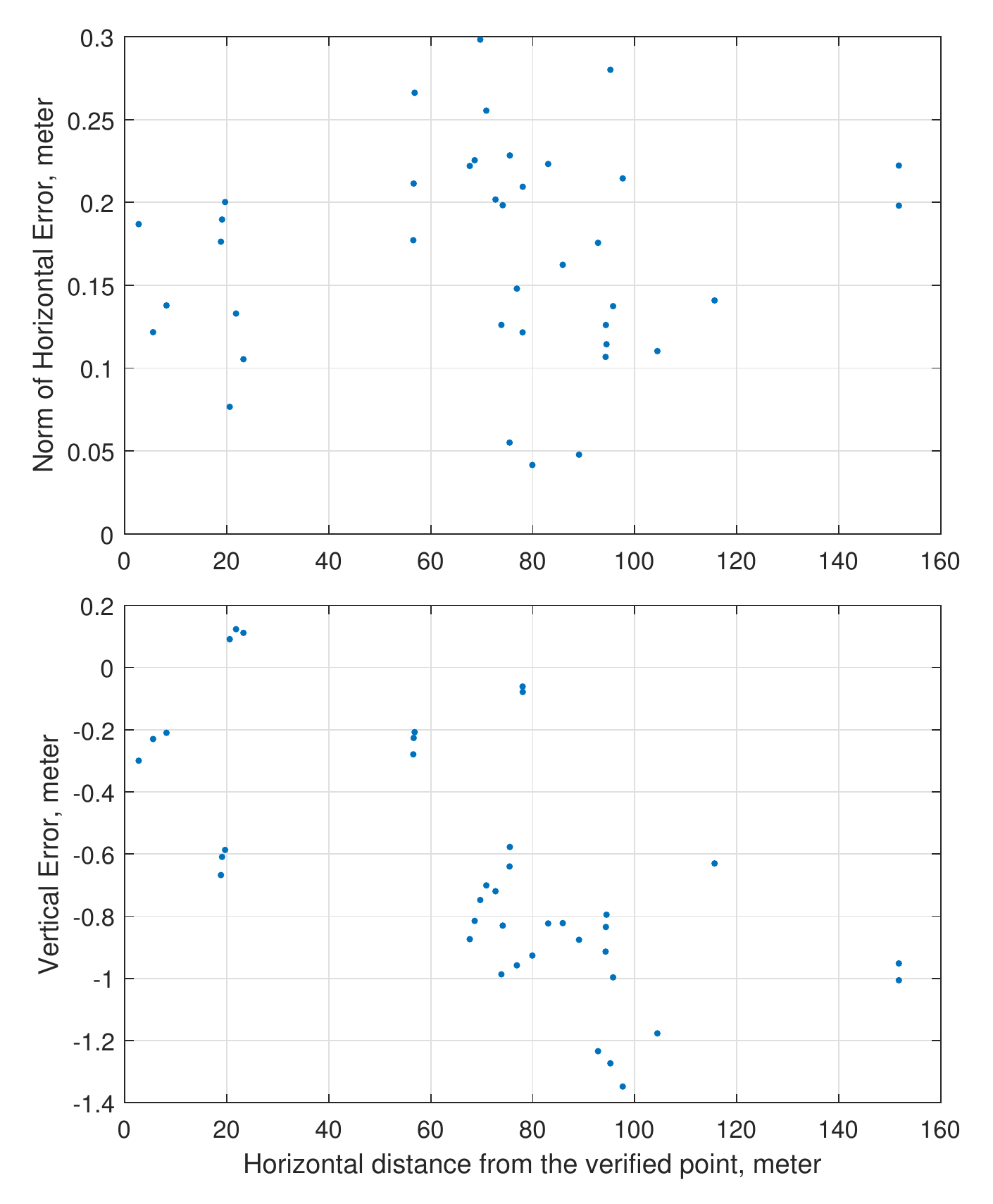}
		\caption{	
			 Horizontal error norm $\delta P_{H_k}^g$ and vertical error $\delta P^g_{D_k}$ \\
			 versus HD to the verified point (i.e., $D_{H_k}$).}
		\label{fig:map1hor}
	\end{subfigure}%
	\begin{subfigure}{.4\textwidth}
		\includegraphics[width=0.9\linewidth]{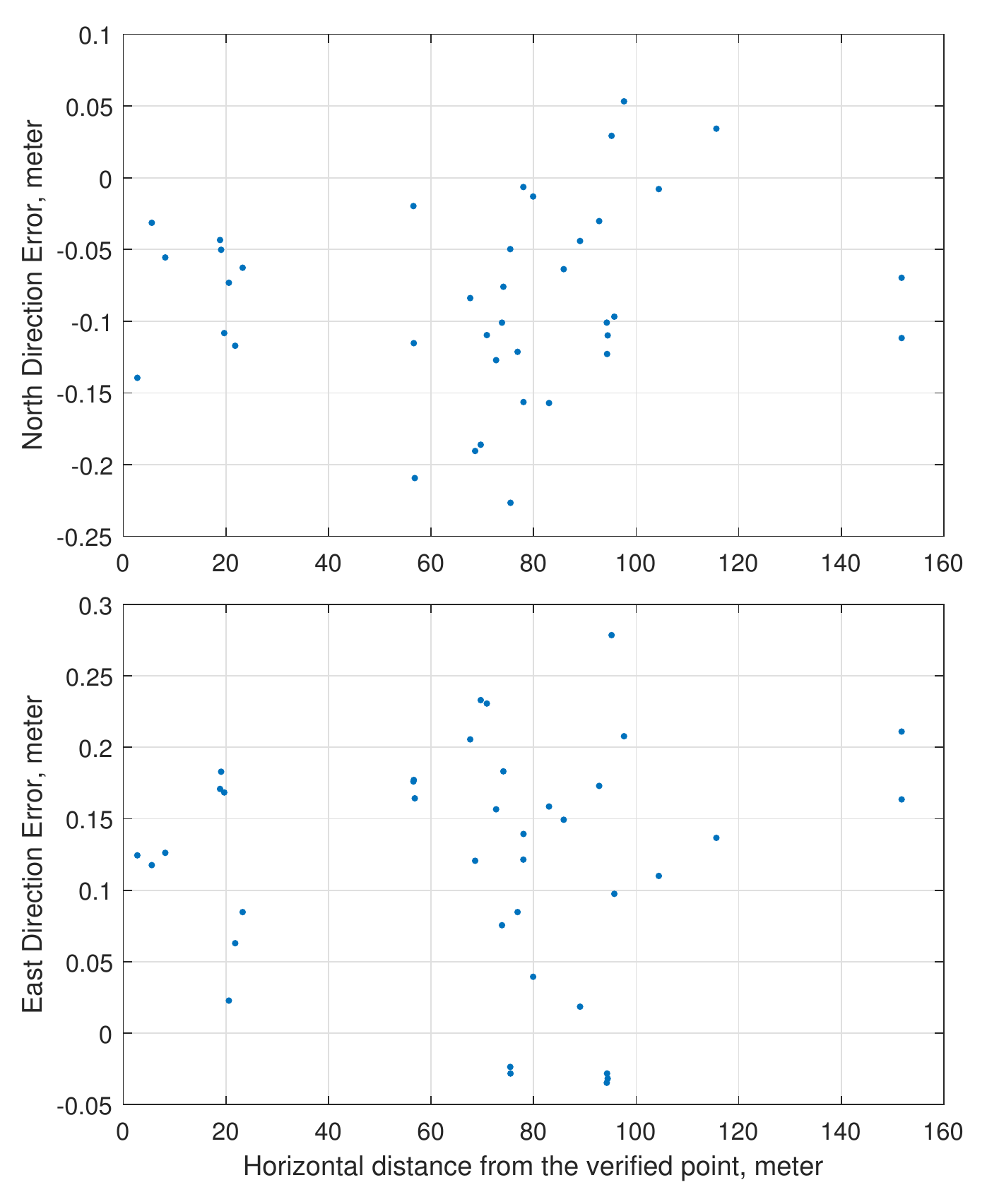}
		\caption{	
			 North and East error components versus \\
			 HD to the verified point (i.e., $D_{H_k}$).}
		\label{fig:map1east}
	\end{subfigure}
	\caption[short]{USDOT map accuracy assessment near UCR CE-CERT. }
	\label{fig:near_Cert}
\end{figure*}

Note that both the north and east components of the position error vector are biased by -0.13 m and 0.21 m, respectively.
Due to the fact that the bias is statistically the same for both feature points (i.e., click tests) and all clicks, this bias is attributed to the error in the placement on the verified point within the USDOT tool. 
See also the discussion of Figs. \ref{fig:map1east} and \ref{fig:map2east}.

\subsection{Feature Mapping Accuracy Analysis}
The USDOT map tool provides geographic coordintes (i.e., Longitude, Latitude, and Altitude) for feature points. 
The geographic coordinates are transferred to \ac{ECEF} coordinates using the method described in Eqns. (2.9-2.11) of \cite{farrell2008aided}, then to local tangent plane using Eqn. (\ref{eqn:Transl_E2G}).
The USDOT location of feature $k$ is denoted as $\hat{\BoldP}^e_k$ and  $\hat{\BoldP}^g_k$.
The GNSS surveyed location of feature $k$ is denoted as ${\BoldP}^e_k$ and ${\BoldP}^g_k$.
The position error for feature $k$ is computed as
	$$\delta \BoldP^g_k = \hat{\BoldP}^g_k - {\BoldP}^g_k = \BoldR^g_e\,(\hat{\BoldP}^e_k-\BoldP^e_k)$$
where $\delta \BoldP^g_k = [\delta P^g_{N_k}, \delta P^g_{E_k}, \delta P^g_{D_k}]^T$ defines the north, east, and down components of the error vector.
The metrics for analyzing the accuracy of the $k$-th feature are
the {\em horizontal error norm}: $$\delta P_{H_k}^g = \sqrt{(\delta P^g_{N_k})^2+(\delta P^g_{E_k})^2};$$ 
and, the {\em vertical error}: $\delta P^g_{D_k}$.
The \ac{HD} between the $k$-th test point and verified point ($\BoldP^g_v=\0$) is
$$D_{H_k} = \sqrt{(\BoldP^g_{N_k})^2+(\BoldP^g_{E_k})^2}.$$

\begin{figure*}[tb]
	\centering
	\begin{subfigure}{.4\textwidth}
		\centering
		\includegraphics[width=0.8\linewidth]{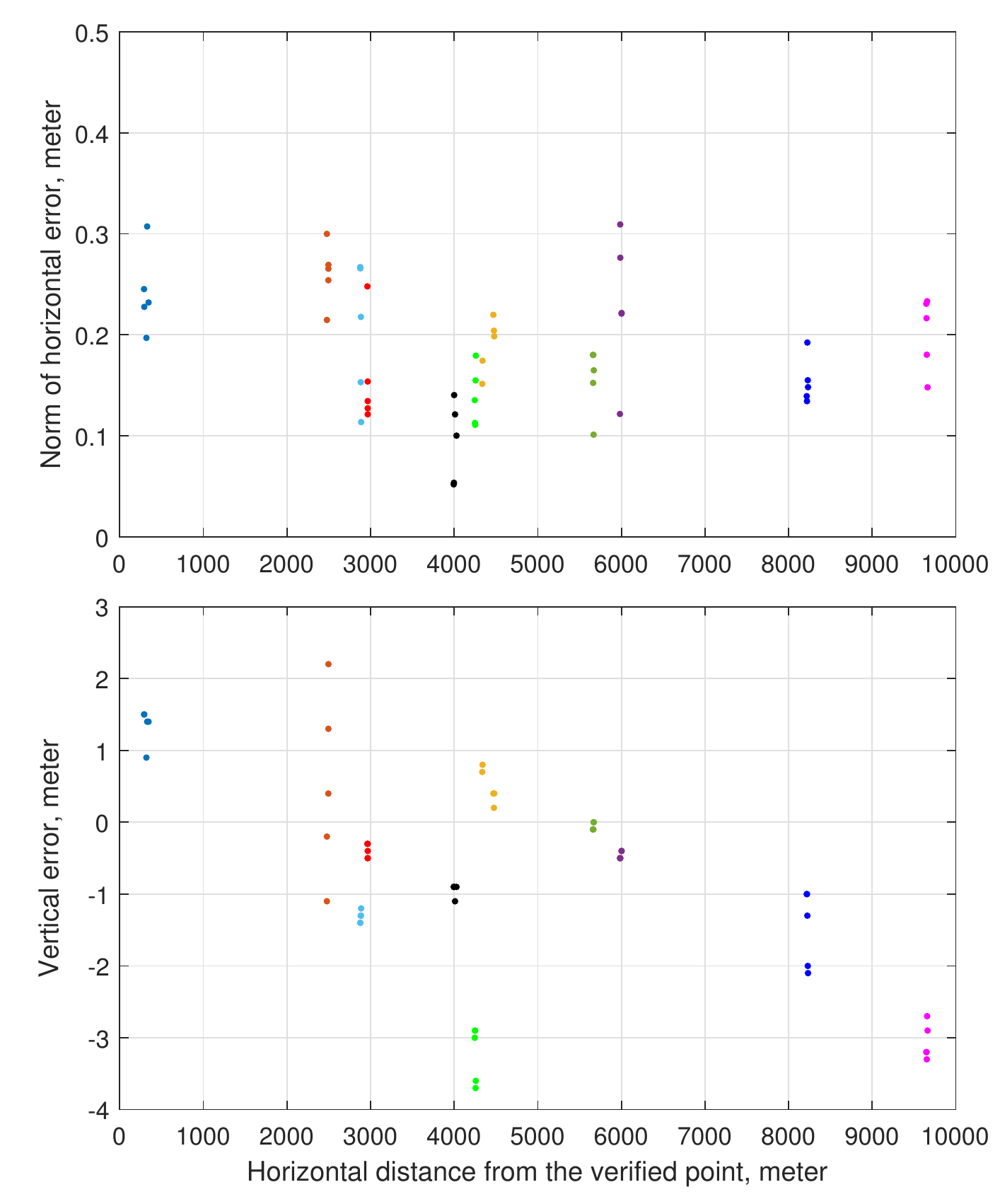}
		\caption{Horizontal error norms and Vertical error}
		\label{fig:map2hor}
	\end{subfigure}%
	\hspace{.05\textwidth}
	\begin{subfigure}{.4\textwidth}
		\centering
		\includegraphics[width=0.9\linewidth]{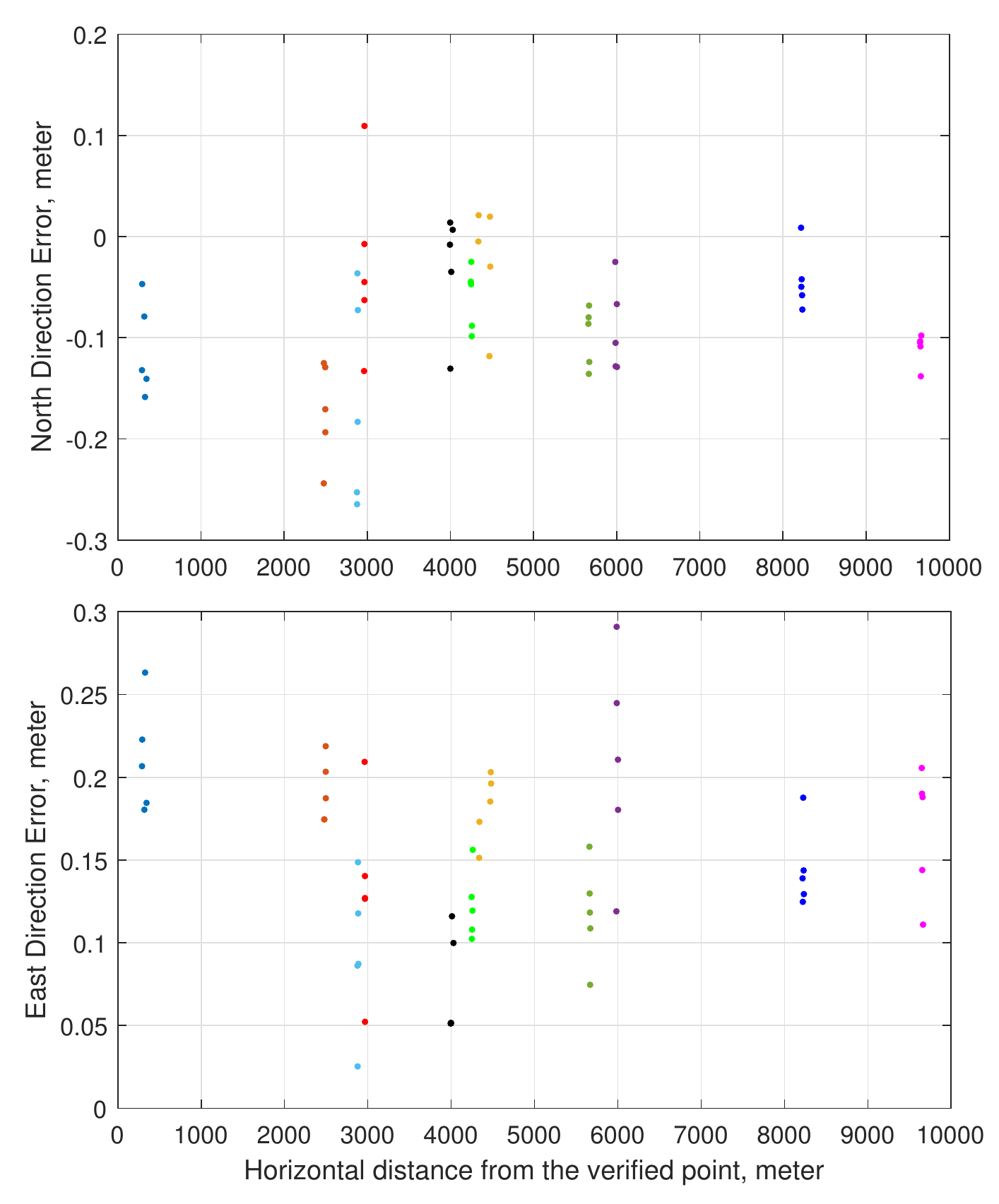}
		\caption{North and East error components}
		\label{fig:map2east}
	\end{subfigure}
	\caption[short]{Graphs of USDOT map accuracy assessment for expanded area. 
		$S_1$ in green with \ac{HD}  near 5.7 km, 
		$S_2$ in dark purple with \ac{HD} near 6 km, 
		$S_3$ in amber with \ac{HD} near 4.4 km, 
		$S_4$ in light blue with \ac{HD} near 2.9 km, 
		$S_5$ in orange with \ac{HD} near 2.4 km, 
		$S_6$ in blue with \ac{HD} near 0.3 km, 
		$S_7$ in black with \ac{HD} near 4 km, 
		$S_8$ in red with \ac{HD} near 3 km, 
		$S_9$ in green with \ac{HD} near 4.2 km, 
		$S_{10}$ in deep blue with \ac{HD} near 8.2 km, and
		$S_{11}$ in magenta with \ac{HD} near 9.7 km.}
	\label{fig:expd}
\end{figure*}

Fig. \ref{fig:near_Cert} displays data for assessing accuracy for the features shown in Fig. \ref{fig:map1} that are near CE-CERT.
Fig. \ref{fig:map1hor} displays the horizontal error norm and vertical error for the feature points near the \ac{UCR} \ac{CE-CERT}.
Fig. \ref{fig:expd} presents  data for the expanded area shown in Fig. \ref{fig:map2}.
The expanded area includes 11 clusters.
Data for each cluster is depicted in a different color in Fig. \ref{fig:expd}.
In each figure the x-axis is the horizontal distance $D_{H_k}$ from the verified point.
Fig. \ref{fig:map1hor} shows 0.17 m mean and 0.30 m maximum horizontal error. 
Fig. \ref{fig:map2hor} shows the horizontal error norm and vertical errors over longer horizontal distances from the verified point. 
Fig. \ref{fig:map2hor} shows 0.18 m mean and 0.31 m maximum horizontal error.
There are no discernible trends in the horizontal error as a function of the distance from the verified point.

Fig. \ref{fig:map2hor} also shows that the vertical error does change as a function of the distance from the verified point. 
The tool georectifies remote sensing satellite imagery to achieve its accuracy in the horizontal directions.
Satellite imagery does not provide depth information; therefore, the underlying vertical accuracy is limited.

Figs. \ref{fig:map1east} and \ref{fig:map2east} show the individual components of the horizontal error. 
In Fig. \ref{fig:map1east} the mean north and east errors are -0.08 m and 0.12 m, respectively.
In Fig. \ref{fig:map2east} the mean north and east errors are -0.08 m and 0.15 m, respectively.
These biases are consistent with each other and with those in Fig. \ref{fig:click}.
This verifies the conclusion that the verified point selected within the USDOT tool is biased by this amount relative to the desired feature point, due to the limited resolution of the imagery in that tool.

The symbol $\sigma_H$ represents the \ac{MSE} of the experimental horizontal position error $\delta P_{H_k}^g$.
 The \ac{MSE} of $\delta P_{H_k}^g$  is 0.18 m for $N_1$ points and 0.20 m for $N_2$ points.
The \ac{MSE} $\sigma_H$ over all 94 feature points is 0.19 m.

\begin{figure}[bht]
	\centering
	\includegraphics[width=0.8\linewidth]{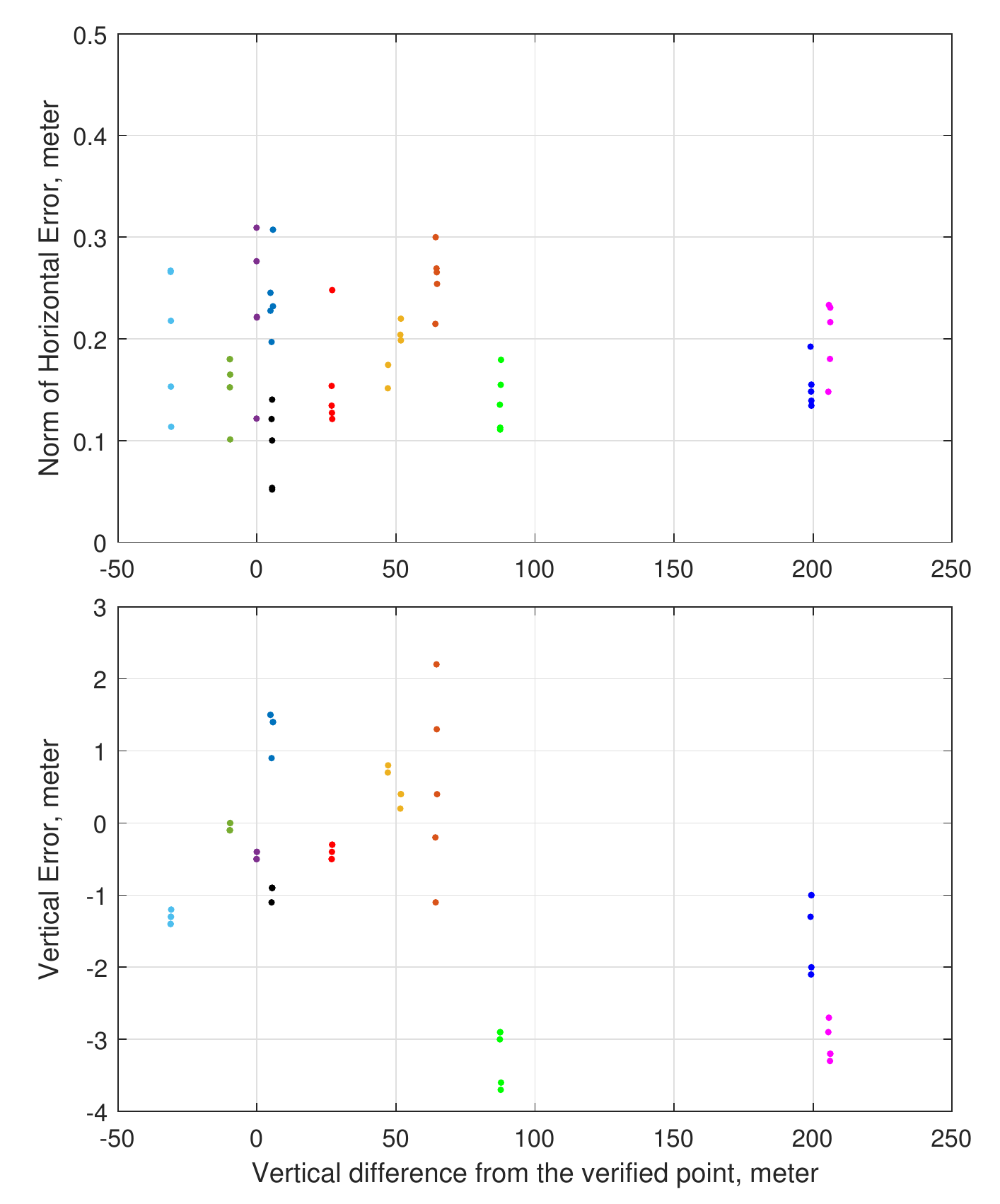}
	\caption{Horizontal error norms and Vertical error of expanded area.}
	\label{fig:map2_vd}
\end{figure}

Fig. \ref{fig:map2_vd} plots the horizontal and vertical errors versus vertical difference relative to the verified point. 
The horizontal accuracy remains constant as elevation changes.
The vertical error is an order of magnitude larger than the horizontal error and does change with both the horizontal and vertical separation from the reference point.

\subsection{Map Horizontal Accuracy Assessment}\label{sect:ass_map}
The experimental horizontal  position error $\sigma_H$ is the result of the four specific errors discussed in Sections \ref{sec:survey} and \ref{sec:mark}, specifically:
$$\sigma_H^2 = 2\,\sigma_C^2 + \sigma_M^2 + \sigma_G^2 + \sigma_S^2.$$
 
The US-DOT click accuracy $\sigma_C$ is multiplied by 2 since it is applied to the clicks for both the feature point and the verified point. 
Since we have experimentally determined values for $\sigma_H$, $\sigma_C$, $\sigma_G$, and $\sigma_S$,
we can compute 
$\sigma_M = \sqrt{ \sigma_H^2 - \left(2 \, \sigma_C^2 + \sigma_G^2 + \sigma_S^2\right)}$. 
Using either value of $\sigma_C$  from the two click tests,
the resulting value of  $\sigma_M$ is $0.17$ m.

	\section{Conclusions and Discussion}
	\ac{HiD} digital maps are an indispensable  automated driving technology for \ac{CAV} applications. 
	The \ac{USDOT} map tool allows users to create MAP and \ac{SPaT} messages with free access, but an assessment of its accuracy does not exist in the current literature.
	This document assessed the accuracy of the US-DOT map tool using a set of 94 feature points with an 10 km area. 
	The assessed mean square horizontal map error is 17 centimeters, which satisfies the lane-level \ac{HiD} map requirement (10-20 cm) \cite{liu2020}. The maximum horizontal map error is less than 30 centimeters.
	The assessment also demonstrated that this horizontal map accuracy was maintained within a 10 km distance of the USDOT map tool verified point that was used in this study.

%

	\black
	\bibliographystyle{bibl/IEEEtran}
	\bibliography{bibl/IEEEabrv,bibl/References}

	\vfill
	\clearpage
\end{document}